\documentclass[12pt,a4paper]{article}
\usepackage{amssymb,amsmath}
\usepackage[dvips]{lscape,graphicx}

\newcommand{\bc}{\begin{center}}
\newcommand{\ec}{\end{center}}
\newcommand{\bd}{\begin{displaymath}}
\newcommand{\ed}{\end{displaymath}}
\newcommand{\be}{\begin{equation}}
\newcommand{\ee}{\end{equation}}
\newcommand{\ba}{\begin{array}}
\newcommand{\ea}{\end{array}}
\newcommand{\bt}{\begin{tabular}}
\newcommand{\et}{\end{tabular}}

\begin{document}
\title{\bf Crypto-baryonic Dark Matter}

\author{C.D.~Froggatt${}^{1,2}$, H.B.~Nielsen${}^{2}$
\\[15mm] \itshape{${}^{1}$ Department of
Physics and Astronomy,}\\[3mm] \itshape{Glasgow University,
Glasgow, Scotland}\\[3mm] \itshape{${}^{2}$ The Niels Bohr
Institute, Copenhagen, Denmark}\\[3mm]
}

\date{}

\maketitle

\begin{abstract}
It is proposed that dark matter could consist of compressed
collections of atoms (or metallic matter) encapsulated into, for
example, 20 cm big pieces of a different phase. The idea is based
on the assumption that there exists at least one other phase of
the vacuum degenerate with the usual one. Apart from the
degeneracy of the phases we only assume Standard Model physics.
The other phase has a Higgs VEV appreciably smaller than in the
usual electroweak vacuum. The balls making up the dark matter are
very difficult to observe directly, but inside dense stars may
expand eating up the star and cause huge explosions (gamma ray
bursts). The ratio of dark matter to ordinary matter is expressed
as a ratio of nuclear binding energies and predicted to be about
5.
\end{abstract}

\newpage

\section{Introduction}

In this note we attempt to obtain a model for dark
matter without introducing any new fundamental particles or
interactions beyond the Standard Model. Our main assumption is
that the cosmological constant is not only fine-tuned for one
vacuum but for several, which we have called \cite{bnp,origin,MPP,bn,fn2}
the Multiple Point Principle (MPP). The existence of another vacuum
could be due to some genuinely new physics, but here we consider
a scenario where it occurs in the pure Standard Model. Indeed,
we have previously speculated \cite{itep,portoroz,coral,pascos04}
that an exotic bound state of 6 top quarks and 6 anti-top quarks
could, due to its surprisingly strong binding via Higgs exchange,
even be on the verge of becoming tachyonic and
form a condensate thereby making up an alternative vacuum.

With the existence of just these 2 degenerate vacua\footnote{With
the added assumption of a third Standard Model phase, having
a Higgs vacuum expectation value of the order of the Planck scale,
we obtained a value of 173 GeV for the top quark mass \cite{fn2}
and even a solution of the hierarchy problem, in the sense of
obtaining a post-diction of the order of magnitude of the ratio of
the weak to the Planck scale \cite{itep,portoroz,coral,pascos04}.}
domain walls would have easily formed, separating the different
vacua occurring in different regions of space, at high enough
temperature in the early Universe. Since we assume the weak scale
physics of the top quark and Higgs fields is responsible for
producing these bound state condensate walls, their energy scale
will be of order the top quark mass. We note that, unlike walls
resulting from the spontaneous breaking of a discrete symmetry,
there is an asymmetry between the two sides of the the wall. So a
wall can readily contract to one side or the other and disappear.

Our main idea is now that dark matter is indeed ordinary baryonic matter or
even atoms packed into small balls - actually they turn out
to be surprisingly big - surrounded
by the walls separating the vacua. In other words we imagine that dark matter
consists of ``small'' particles which are ordinary atoms in a tiny bit of
the other type of vacuum. The Higgs vacuum expectation value will
be of the same order of magnitude in the alternative phase
and we assume it is reduced by, say, a
factor of two. This would reduce the quark masses by a
factor of two also and the pion mass by the square root of two,
in going to the alternative phase reigning inside the dark matter balls.
This, in turn, would make the range of the nuclear force longer,
causing a stronger binding of the nuclei by an amount comparable
to the binding they already have in normal matter.

A major problem for making all the dark matter out of normal baryonic
matter is, of course, that it would spoil the successful big bang
nucleosynthesis (BBN) calculations. However, this problem is
avoided in our model by having the dark matter nucleons
encapsulated by the walls, where they would be relatively inert.
We should note that a model for dark matter using an alternative
phase in QCD has been proposed by Oaknin and Zhitnitsky \cite{OZ}.

\section{Formation of Dark Matter}

We should now look if we can
find a scenario for how the dark matter balls could be formed:

Let us denote the order parameter field describing the new bound state
which condenses in the alternative phase by $\phi_{NBS}$. It would
fluctuate statistical mechanically and, as the temperature T in the
early Universe fell through the weak energy scale, the expected
distribution of the $\phi_{NBS}$-field
\begin{equation}
P(\phi_{NBS}) d\phi_{NBS} = A \exp(-\frac{V_{eff}(\phi_{NBS})}{T^4})
d\phi_{NBS}
\label{fielddist}
\end{equation}
would have become more and more concentrated around the - assumed
equally deep - minima of the effective potential
$V_{eff}(\phi_{NBS})$.

If we could trust the estimate from formula (\ref{fielddist})
the wall density would go down exponentially, being proportional
to $\exp(-\frac{V_{max}}{T^4} O(1))$, where $V_{max}$ is the
maximum value of the effective potential. This would happen
if the Hubble
expansion were adiabatically slow, but that is not realistic.
There is at least the Kibble mechanism \cite{kibble}
ensuring that the distances between the walls will not become longer than
the horizon distance, {\em provided in practice there is an effective symmetry
between the phases}.
This is due to the two vacua appearing with about equal probabilities
in regions separated by an horizon distance.
There was indeed an effective symmetry between the vacua as
the temperature fell below the energy scale of the walls, since the
vacua had approximately the same free energy densities.
Eventually the small asymmetry between these
free energy densities would have led
to the dominance of one specific phase inside each horizon region and,
finally, the walls would have contracted away.
However it is a very detailed dynamical question as to how far
below the weak scale the walls would survive. It seems quite possible
that they persisted until the temperature of the Universe fell
to around 1 MeV.

In our favourite scenario we imagine that the disappearance of the
walls in our phase - except for very small balls of the fossil
phase -  occurred at the time when the temperature $T$ was
of the order of 1 MeV to 10 MeV. At this epoch, the
nuclear forces and the mass difference between the nucleons
inside and outside the balls became relevant.
For instance we expect the effect
of the mass difference $\Delta m_N$ between the nucleons in
the two phases to lead to a nucleon density ratio of
$\exp(-\Delta m_N/T)$.
Using an additive quark mass dependence approximation for the
nucleons \cite{weinberg} and a Higgs VEV reduced by a factor of 2
in the alternative phase, we obtain a
difference between the nucleon masses in our phase and in the alternative one
of the order of 10 MeV. Thus, as the temperature fell below 10 MeV,
the nucleons collected more and more strongly into
the alternative phase.

However as the alternative phase bubbles contracted and the bubble radii
got smaller, the chance for a nucleon that happened to be in our phase
hitting a piece of alternative phase was reduced.
It is important for our model to estimate the critical nucleon density
inside the alternative phase balls at which the collection of nucleons into
them stopped.  This is because any nucleons that might be expelled
from the balls, after the density had increased above this critical one,
would {\em no} longer be reabsorbed by the balls. Such nucleons would have
to stay forever outside the balls and make up {\em normal matter}.

Let us define a parameter $\Xi$ as the ratio of the density of walls
compared to the density as would be ensured by the Kibble mechanism
(which means wall-distances equal to the horizon length) at
that time when the wall contraction started being rapid, and we
imagine the proper balls began to form.
The nucleon velocity $v$ multiplied by $\Xi$
gave the probability during a Hubble time for a
nucleon in our phase to hit a piece of the alternative phase.
We define the ball radius $r$
in units of the starting radius:
\begin{equation}
r = ``radius\mbox{''} /(``horizon\mbox{''} /\Xi)
\end{equation}
where $``radius\mbox{''}$ and $``horizon\mbox{''}$ are the radius and
horizon distance
when the contraction
started to gain speed. Then the collection of nucleons into the alternative
phase stopped being effective when  $ r^2 \Xi v$ became of order unity.

If this density increase was on a shorter time scale than the
Hubble scale, the temperature remained essentially the same during the
contraction. Due to the  increase in density, however,
successively heavier and heavier nuclei could form.

Let us consider
the formula \cite{kolbturner} for the temperature $T_{NUC}$
at which a given species of
nucleus with nucleon number A can become copious from pure
statistical mechanics, ignoring Coulomb repulsion:
\begin{equation}
T_{NUC} = \frac{B_A /(A-1)}{\ln(\eta^{-1}) + 1.5 \ln(m_N /T_{NUC})}.
\end{equation}
Here $B_A$ is the binding energy of the nucleus - in the phase in
question of course - $\eta$= $\frac{n_B}{n_{\gamma}}$ is the ratio of
the baryon number density relative to the photon density, and
$m_N$ is the nucleon mass. In our phase, for example, the
temperature for ${^4}$He to be thermodynamically favoured turns out
from this formula to be 0.28 MeV. In the other phase, where the
Higgs field has a lower VEV by a factor of order unity,
the binding energy $ B_A$ will become bigger by a factor
of order unity.

We speculate that the formation of the light nuclei up to helium,
and indeed mainly ${^4}$He, occurred before
the collection of the nucleons into the balls stopped, while the
heavier nuclei first formed after the density had increased
further and the collection had stopped.

We must now discuss if we can imagine a value of our parameter $\Xi$
such that our scenario can function:

a) Since the Kibble mechanism would take over and ensure that there be
at least one wall met per horizon distance, we must have $\Xi \ge 1$.

b) We would like the helium four to have formed before the collection
mechanism putting the nucleons into the alternative phase stopped.
This occurred when $1 \approx r^2 \Xi v$ for a nucleon velocity $v$.
We could achieve the thermodynamical equilibrium
point for this fusion at 1 MeV with an $\eta$ of order say $10^{-3}$ rather
than the $10^{-9}$ or $10^{-8}$ which would be there without the
walls. Such a concentration would correspond to $r=10^{-2}$.
That is to say then that we need $\Xi > 10^5$, where we
use the estimate $v=10^{-1}$.

c) On the other hand we would like the fusion into the heavier
nuclei such as carbon first to occur {\em after} the switch off of
the collection of the nucleons. Since the equilibrium temperature
$T_{NUC}$ will not be very different for the formation of helium or
the higher nuclei, we have to rely on the
Coulomb barrier to prevent the two types of fusion going
on at the same time. However it is not easy to estimate how much
smaller the radius of the ball would be, when finally the further
fusion took place. A reduction by at least an order of magnitude
would seem reasonable, so that $\Xi$  is allowed in the
region up to $\Xi = 10^6$ but could possibly be higher.

d) On the other hand we also need that the fusion to higher
elements did not come too late or not at all.
We need it to occur before the temperature reached about 1 MeV,
so as not to disturb the BBN in our phase.
There is the following hope: When the balls become very
concentrated, the density could rise up to near one nucleon per volume
of a sphere with a radius say of the order $\frac{1}{10}$ MeV$^{-1}$.
At a temperature of 1 MeV there would then only be a negligible amount
of positrons and photons compared to the nucleons. If,
under these conditions, a fusion process locally liberates an
energy of 1.4 MeV per nucleon, or say 0.5 MeV per degree of freedom,
the temperature would rise by an amount
comparable to the prevailing temperature. Thus
provided the density had reached $\eta > 1$, so that the
baryons dominated, fusion processes could have been
triggered off by such a temperature increase.
There was then the possibility of a chain reaction and
an explosive heating of the whole ball, provided
$\eta >1$ and thus $r < 10^{-3}$ at the time of the helium
to heavy nuclei burning.
The requirement $r^2 \Xi v < 1$, ensuring that
the recollection of the nucleons into
the alternative phase would have stopped by then, implies that
$\Xi < 10^7$.

From these considerations, we obtain the suggested range
$10^5 < \Xi < 10^7$
for the parameter $\Xi$.

The energy set free by the fusion from helium into the heavier nuclei,
such as carbon or iron, went
into raising the temperature of the motion of the nuclei
or into nucleons that were evaporated out of the
nuclei. Provided the number of free nucleons present inside the balls
before the fusion of the helium into heavier nuclei was much
smaller than the number evaporated,
the chemical potential and temperature in the balls would soon
have reached
the level where free nucleons would spill over the
wall to the outside of the ball.
Most of the evaporated nucleons
would be formed after this spill over temperature was reached and
just run out of the ball, forming the normal matter which
underwent the usual BBN in our phase.

From a simple energy conservation argument,
we can now obtain the ratio of evaporated free nucleons
relative to the remaining nucleons, which are now inside the
rather heavy nuclei formed by the internal fusion.
Inside the ball, both
the free nucleons and the nucleons inside the nuclei had an extra
amount of energy due to the chemical potential and the
temperature. We take these extra
amounts of energy to be the same, whether the nucleons were inside
or outside the nuclei. This means that we can simply estimate the
amount of energy
released by the fusion as if it occurred in the inside  vacuum
and without any appreciable amount of nucleons around.

In the ${^4}$He nucleus,
the nucleons have a binding energy of 7.1 MeV in normal matter in
our phase, while a typical ``heavy'' nucleus has a binding energy
of 8.5 MeV for each nucleon \cite{ring}.
Let us, for simplicity, assume that the ratio of these
two binding energies per nucleon is the same in the alternative
phase and
use the normal binding energies in our estimate below.
Thus we take the energy released by the fusion of the helium
into heavier nuclei to be 8.5 MeV - 7.1 MeV = 1.4 MeV per nucleon.
Now we can calculate
what fraction of the nucleons, counted as {\it a priori} initially
sitting in the heavy nuclei, can be released by this
1.4 MeV per nucleon. Since they were bound inside the nuclei
by 8.5 MeV relative to the energy they would have outside,
the fraction released should be
(1.4 MeV)/(8.5 MeV) = $0.16_5$ = 1/6.
So we predict that the normal baryonic matter
should make up 1/6 of the total amount of matter, dark as well as
normal baryonic. This is in agreement with
astrophysical fits \cite{spergel}, which give
the amount of
normal baryonic matter relative to the total matter to be
$\frac{4\%}{23\% + 4\%} = 4/27 = 0.15$.


Let us admit that
a slightly different scenario is possible:
Provided the balls are sufficiently
big they would have functioned as just phases far away from
the region where the
BBN went on and would therefore not have disturbed it.
If one phase did not have time to collect almost all the nucleons,
the ratio of dark to normal matter would not be predicted, but would
naturally become of order unity. This in itself is a remarkable
result.

\section{ Properties of Dark Matter Balls.}

The size of the balls are not
safely predicted in our model and we should
rather use the parameter $\Xi$ to parameterize the
mass of the balls and thereby also their number density.
Their size also depends sensitively on the order of magnitude
assumed for the wall energy density.
Fits to BBN suggest that there be about $10^{-9}$ baryons per photon.
Thus there were of the order of  $10^{54}$
baryons in the horizon region at a temperature of 1 MeV and time scale of 1 s,
when we defined the $\Xi$ parameter. This means that we expect
of the order of $\Xi^3$ balls to have formed per horizon
volume with its $10^{54}$ baryons. So the number of baryons
$N_B$ in each ball is of the order
\begin{equation}
N_B = 10^{54}/\Xi^3.
\label{NB}
\end{equation}

We now consider the stability condition for the balls. For a ball
of radius R the ``weak scale'' tension $s \approx (100\ \mathrm{GeV})^3$
of the wall provides a pressure $s/R$. The energy needed to release
a nucleon from the alternative vacuum into our vacuum is
approximately 10 MeV. So the maximum value for the electron Fermi
level inside the balls is $\sim 10$ MeV, since otherwise it would
pay for electrons and associated protons to leave the alternative
vacuum. In order that the pressure from the wall should not quench
the corresponding maximal electron pressure of (10 MeV)$^4$, we
need $s/R < (10$ MeV$)^4$, which means $R > R_{crit} = 2\ \mathrm{cm}$. If
the balls have a radius smaller than $R_{crit}$, they will implode.
These critical size balls have a nucleon number density of
(10 MeV)$^3$ and thus contain of order $N_B = 10^{36}$ baryons and
electrons. It follows from Eq. \ref{NB} that $\Xi_{crit} = 10^6$
and thus ball stability requires $\Xi < 10^6$, which restricts our
allowed range of $\Xi$ further to $ 10^5 < \Xi < 10^6$.

Let us therefore consider a typical ball as corresponding to $\Xi
= 3\times 10^5$, which has a radius of order 20 cm. It has an
electron Fermi momentum of order 5 MeV and contains of order $N_B
= 3\times 10^{37}$ baryons and has a mass of order $M_B = 10^{11}$
kg = $10^{-19}M_{\odot} = 10^{-14}M_{\oplus}$.
Therefore dark matter balls can not be revealed by microlensing
searches, which are only sensitive to massive astrophysical
compact objects with masses greater than
$10^{-7}M_{\odot}$ \cite{afonso}.
Since the dark matter density is 23\% of the critical density
$\rho_{\mathrm{crit}} = 10^{-26}$ kg/m$^3$, a
volume of about $10^{37}$ m$^3$ = (20 astronomical units)$^3$
will contain on the average just one dark matter ball.

Assuming the sun moves with a velocity of 100 km/s relative to the
dark matter and an enhanced density of dark matter in the galaxy of
order $10^5$ higher than the average, the sun would hit of order
$10^8$ dark matter balls of total mass $10^{19}$ kg in the lifetime
of the Universe. A dark matter
ball passing through the sun would plough through a mass of sun
material similar to its own mass. It could therefore easily become
bound into an orbit say or possibly captured inside the sun, but be
undetectable from the earth. On the other
hand, heavy stars would tend to capture most of the dark matter
balls impinging on them. However the $10^4$ or so dark matter balls
hitting the earth in the lifetime of the Universe would go through
the earth without getting stopped appreciably.

It follows that DAMA \cite{dama} would not have any chance of seeing
our dark matter balls, despite their claim to have detected a
signal for dark matter in the galactic halo.
However EDELWEISS \cite{edelweiss}, CRESST
\cite{cresst} and CDMS \cite{cdms} do not confirm the effect seen by
DAMA. It is also possible that DAMA saw something other than dark
matter. Geophysical evidence for the dark matter balls
having passed through the earth would also be extremely difficult to find.

In principle the balls are only metastable and will at the end
explode, by the wall tunnelling away leaving the material outside.
However, in reality, they are very stable and it is highly
unlikely that a given ball could be close enough to the stability
border for an explosion to actually take place. On the other hand,
we could imagine that dark matter balls had collected into the
interior of a collapsing star. Then, when the density in the
interior of the star gets sufficiently big, the balls could be so
much disturbed that they would explode. The walls may then start
expanding into the dense material in the star, converting part of
the star to dark matter. As the wall expands the pressure from the
surface tension diminishes and lower and lower stellar density
will be sufficient for the wall to be driven further out through
the star material. This could lead to releasing energy of the
order of 10 MeV per nucleon in the star, which corresponds to of
the order of one percent of the Einstein energy of the star!
Such events would give rise to really huge energy releases,
perhaps causing supernovae to explode and producing the canonballs
suggested by Dar and De Rujula \cite{deRujula} to be
responsible for the cosmic gamma ray bursts. We should note that a
different (SUSY) phase transition inside the star has already been
suggested \cite{clavelli} as an explanation for gamma ray bursts.

A dark matter ball can also explode due to the implosion of
its wall.
Such an implosive instability might provide a mechanism for
producing ultra high energy cosmic rays from seemingly empty
places in the Universe. This could help resolve the
Greisen-Zatsepin-Kuzmin \cite{G,ZK} cut-off problem.

\section*{Acknowledgements}

We acknowledge discussions with D.~Bennett and Y.~Nagatani in the
early stages of this work. CDF would like to acknowledge the hospitality
of the Niels Bohr Institute and support from the Niels Bohr Institute
Fund and PPARC.

\end{document}